\documentclass[cits]{PoS}

\usepackage{subfigure}
\usepackage{wrapfig}
\usepackage{booktabs,amsmath}
\usepackage{macrosn,lat13}

\title{Scale setting in lattice QCD}

\ShortTitle{Scale setting in lattice QCD}

\newcommand{\preprintline}{\vspace{3cm}\newline
\rightline{\parbox{2.9cm}{\large\tt DESY 13-262}}
}

\author{Rainer Sommer\\
        NIC @ DESY, Platanenallee 6, 15738 Zeuthen, Germany\\
        E-mail: \email{Rainer.Sommer@desy.de}}

\abstract{The principles of scale setting 
in lattice QCD as well as the advantages and disadvantages of
various commonly used scales are discussed. After listing criteria
for good scales, I concentrate on the main presently used ones
with an emphasis on scales derived from the 
Yang-Mills gradient flow. For
these I discuss discretisation errors,
statistical precision and mass effects. A short review on
numerical results also brings me to an unpleasant disagreement 
which remains to be explained.
\preprintline}

\FullConference{31st International Symposium on Lattice Field Theory - LATTICE 2013\\
		July 29 - August 3, 2013\\
		Mainz, Germany}

\begin{document}

\section{Introduction}
Presumably the most natural scale for low energy QCD is 
the mass of the proton, $\mprot$. It is very well knwown.
Alternatively, as a theorist, one might 
like the pseudo scalar decay constant in the chiral limit, called $f$. 
It sets the scale for interactions of the very low
energy description of QCD.
Indeed, 
let us consider for a little while just QCD in the chiral limit, 
in order to simplify the discussion. 
In this limit, ratios of pairs of these scales or any
other observables of dimension mass, $m_i$,  are predictions of the
theory, but not the scale itself. The latter, and only it, has to be taken from experiment.
In lattice QCD this manifests itself in the fact that 
dimensionless ratios,
\bes
    \Rho_i = m_i / \mprot
\ees
can be computed and {\em have a continuum limit}. Taking 
$\mprot$ as a reference here is what we call scale setting.
Equivalently, a lattice computation determines dimensionless
quantities $M_i=m_i a,\; \Mprot=\mprot a$ as a function of the
bare coupling $g_0$. The prediction for the physical, dimensionful quantity, $m_i$ is then obtained by
\bes
  m_i = \Rho_i \mprot \,,\quad \Rho_i = 
    \lim_{\Mprot\to0} \frac{M_i}{\Mprot} 
  = \lim_{g_0\to 0} \frac{M_i(g_0)}{\Mprot(g_0)} \,.
\ees
In lattice slang, scale setting usually is refered to the 
equivalent point of view of determining the lattice spacing $a$
at a given $g_0$ from the specific quantity $\mprot$,
\bes
    a_\prot(g_0) = \frac{\Mprot(g_0)}{\mprot}\,,
\ees
with the purpose of then defining any other dimensionful
quantity 
\bes
  \mlat_i(g_0) = \frac{M_i(g_0)}{a_\prot(g_0)} \,.
\ees
Clearly $\mlat_i(g_0)$ has a continuum limit and 
at finite $g_0$ (finite $a$) it has lattice artifacts 
(deviations from this limit)
which depend also on the choice of scale, $\mprot$. 

In practise, we also approximate QCD by an effective 
theory where heavy quarks are removed (one often says integrated out).
How good this approximation is also depends on the
choice of the scale as is apparent from the previous discussion.

\subsection{Scale setting is important}

Why is scale setting discussed in a plenary talk?
Both in planning simulations and in the analysis
of the results, the scale (as explained above, whether I say the 
scale or the lattice spacing is just the same) is usually very important.
When quark masses are neglected, an error made in the
scale, systematic or statistical, propagates linearly into a 
hadron mass. 

Now thinking about the real theory, the one with quark masses, 
the scale already enters decisively into fixing the 
bare quark masses in the Lagarangian, namely in planning the simulations.
Usually we do this by adjusting the pseudo scalar meson masses
to their physical values, since those are most sensitive to 
quark masses. The scale is needed for that. It is very important
to perform simulations at or close to the right quark mass, or 
on a given desired trajectory running through the physical point, 
in the parameter space given by the 
quark masses. For mass-degenerate u,d quarks, a  
trajectory may be given by fixed 
(physical or bare PCAC) strange 
(and ultimately charm quark) masses or for example
by a fixed trace of the quark mass matrix,  $\tr m_\quark = \rm const$.
In the past several collaborations missed the proper trajectory by more 
than what they would have liked to.

Furthermore, the scale enters where a momentum in a form factor
is fixed or into the overall size of the lattice. 
It is very useful to know the scale beforehand.

We should also draw lessons from recent years.  In a few cases, challenging
new computations have been carried out, relying on the scale from earlier 
work. Later it turned out that there was an inaccuracy in the scale,
which significantly influenced the result. Examples are changes
in the decay constant $\fds$ of the HPQCD collaboration
due to a reanalysis of $r_1$ \cite{Follana:2007uv,Davies:2010ip} 
and one that the ALPHA collaboration made, namely 
a roughly 20\% change in the two-flavour $\Lambda$-parameter 
after we had computed $r_0/a$ rather than taking it from the 
literature \cite{QCDSF:r0}.

All of this calls for care in the selection and computation of the 
scale.

\subsection{What is a good scale?}
Unfortunately, the proton mass is not easily determined with good precision 
in lattice QCD computations due to a 
large noise/signal ratio in the proton correlation function,
see \sect{s:momega}.  The chiral scale $f$ is related to
experiments only through the chiral perturbation theory 
expansion. Consequently it is not common to use these
observables to set the scale. 
In general we should search for a quantity which\\  
\hspace*{5mm}\onecol{10cm}{\vspace*{-1mm}
\bi \setlength{\itemsep}{0pt}
\item[(P1)]
is computable with a low numerical effort, 
\item[(P2)] 
has a good statistical precision,
\item[(P3)]
has small systematic uncertainties and
\item[(P4)]
has a weak quark mass dependence.
\ei
}
\\
The first two properties are self explanatory, but 
the others require some details. By a good 
systematic precision we mean first of all that the
systematic error in the determination of the numbers
$M_i=a m_i$ for given bare coupling and quark masses is small.
For example such a systematic error may come from 
finite size effects or the contamination by excited states.
A second systematic uncertainty is the discretisation error. 
This is not easily judged. We will comment on it as we go along.
Concerning the quark masses, it is of course useful
to have a scale which depends weakly on them as this 
makes the tuning of the quark masses rather independent 
from the scale and, when the simulations are not at the physical point,
the extrapolation/interpolation to it is easier. Note that the
sensitivity to quark masses also depends on the trajectory 
one chooses to reach the physical point. For example the
mass of the Omega-baryon has a weak dependence on the light
quark masses when the strange quark mass is fixed, but not when
one is on the trajectory  $\tr m_\quark = \rm const$. On such a 
trajectory an appropriate average baryon mass has a weak quark-mass
dependence, see e.g. \cite{Cooke:2013qqa}.

For later use I define a measure 
for the quark mass dependence, 
\bes
  \sm_{Q} = \frac{Q|_{\mpi=500~\MeV} - Q|_{\mpi=130~\MeV}}
      {Q|_{\mpi=130~\MeV}}\,,
      \label{e:sm}
\ees
where $Q$ labels different scales, e.g. $Q=m_\prot$, and 
we assume that we are on a trajectory with strange and charm (bare, PCAC) mass 
fixed. Selfconsistently, MeV come from the scale 
$Q$. Since there is usually a rather linear dependence of
scales on $\mpi^2$, one has 
$\sm_Q \approx \frac{0.23~\GeV^2}{Q|_{\mpi=130~\MeV}}\frac{\partial Q}{\partial_{\mpi^2}}$. 

The statistical precision is determined both by the 
integrated autocorrelation time and by the variance.\footnote{For the definition of variance and autocorrelation
times for derived observables (such as masses extracted from a correlation
function, the primary observable) see \cite{UWerr}.}
These depend on the update algorithm and on the
chosen estimators, respectively. 
E.g. in the simplest case, using or not-using or stochastically-using
translation invariance are different estimators of a correlation function. 


After this preparation, let me come to a discussion of a few 
scales which are in frequent use or newly proposed. I differentiate
between phenomenological scales
and theory scales. The former are related to physical
observables through a minimum amount of theory, while
the latter are constructed to be well computable in lattice
QCD but their values can (at present) only be 
computed in lattice QCD using a phenomenological scale
as an input. Thus a phenomenological scale is
needed in any case. The distinction between one category and the other 
is not sharp. For example I place $r_0$ \cite{pot:r0} with the 
theory scales, although (vague) phenomenological considerations
led to the prediction $r_0\approx 0.49$fm which is not far 
from our present knowledge.

\section{Phenomenological scales}
\subsection{The mass of the $\Omega$-baryon.}
\label{s:momega}
\begin{figure}[ht!]
\centering
   \includegraphics[width=0.9\textwidth]{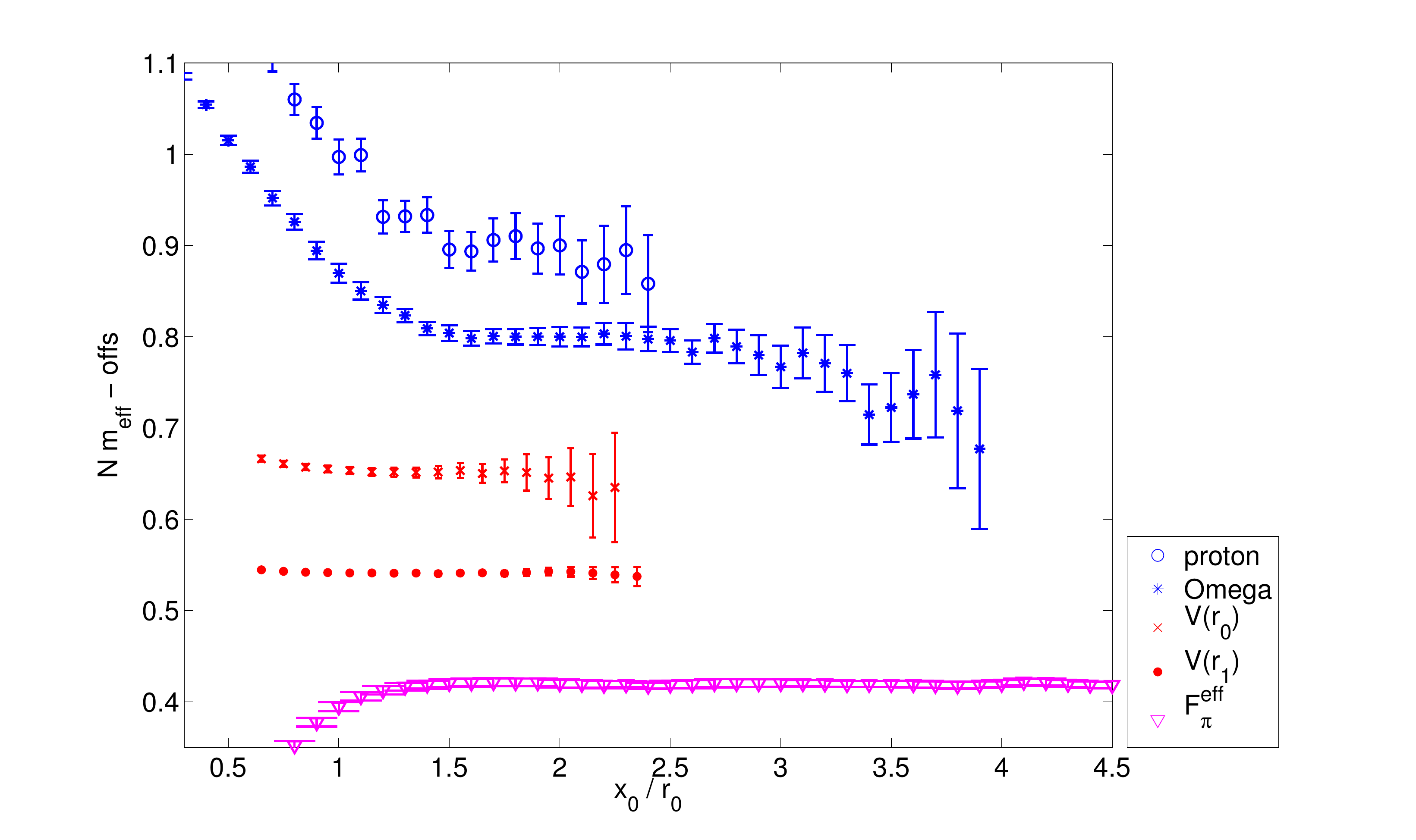}
   \vspace*{-2mm}
\caption{Effective masses for 
$m_\prot$ \cite{lat13:mainz},
$m_\Omega$ \cite{momega:cls},
$V(\approx r_0)$, $V(\approx r_1)$~\cite{alpha:lambdanf2}
and $\fpi$~\cite{lat13:Stefano} on CLS ensemble N6 (see \cite{lat13:Stefano}).
All effective ``masses'' have been scaled 
such that the errors in the graph reflect
directly the errors of the determined scales. They have been shifted vertically.
\label{f:plateaux}
}
\end{figure}

The relative errors $R_{N/S}$ of baryon correlation functions 
grow at large time, $x_0$, as \cite{sn:lepage,LH:martin}\footnote{Note that 
the arguments of \cite{sn:lepage,LH:martin} refer to the 
square root of the variance 
of the correlation functions; a $x_0$-dependence of the 
autocorrelation time is negelceted. 
}
\bes
  R_{N/S}^\prot &\simas{x_0\; \mathrm{large}}& K_\prot\, \exp((\mprot-\frac32\mpi)\,x_0) \approx \exp(x_0 / 0.27\,\fm) \,,
  \nonumber \\[-1.5ex] \label{e:rns} \\[-1.5ex] \nonumber
  R_{N/S}^\Omega &\simas{x_0\; \mathrm{large}}& K_\Omega\,\exp((m_\Omega-\frac32\metas)\,x_0) \approx 
     \exp(x_0 / 0.31\,\fm)  \,,
\ees
using $\metas^2\approx 2\mk^2-\mpi^2$.
The difference in the two formulae is one reason, why there is 
hardly a computation where the scale is taken from the proton mass,
but a few groups like to set it through $m_\Omega$.
The numerical values inserted for the masses are taken
from the PDG and refers to the physical point. Unfortunately it is not 
common any more to publish a plot of the effective mass to illustrate the quality of the plateau. In \fig{f:plateaux} I show one using data from CLS ensemble N6, with $a=0.045\,\fm, \mpi=340\,\MeV$ on a $48^3 \, 96$ lattice
(for details on these ensembles see \cite{lat13:Stefano}),  
which has a statistics of about 8000~MDU $\approx 40 \tau_\mathrm{exp}$.
One observes smaller statistical errors
compared to the proton, but in fact the relation to 
the large time asymptotics, \eq{e:rns},
is not entirely obvious. The plateaux for $m_\Omega,m_\prot$ appear to start
around 0.8~fm$\approx1.6\,r_0$ or so. Somewhat earlier plateaux (at $\approx 0.6\,\fm$) 
are visible in
\cite{Arthur:2012opa,Aoki:2009ix}, where correlation functions
with different smeared 
interpolating fields are considered which use
gauge fixing. 

A second advantage of $m_\Omega$ is the weak dependence
on the light quark mass, if the strange mass is fixed. 
Reversely, it depends much more on the strange mass than other
common choices.

\subsection{Pseudo scalar decay constants $\fpi,\fk$}

Pseudo scalar decay constants have been popular with several collaborations.
In particular they have been used to perform the primary scale setting and then determine the values of the 
theory scales $r_0,r_1,t_0,w_0$ in terms of these. 
A drawback of decay constants is that experimentally they are 
determined from weak processes. The $\pi \to \ell \nu$ decay rate
yields the product $\vud\fpi$ and the decay rate $K \to \ell \nu$ 
is given by $\vus\fk$. Thus the precision we can achieve for 
$\fpi,\fk$ is limited by our knowledge
of the CKM matrix elements $\vud$ and $\vus$, where in particular 
one needs to assume a dominance by the standard model processes and 
a correct determination of the matrix elements of other processes from
which  $\vud$ and $\vus$ are derived.

On the other hand, a clear advantage
is the small and almost $x_0$-independent variance of the 
pseudo-scalar correlators, leading to long plateaux, see \fig{f:plateaux}. 
I will come back to the importance of long plateaux in the 
conclusions.

\subsubsection{Autocorrelations}
\begin{wrapfigure}{l}{6.5cm}
   \vspace*{-5mm}
   \hspace*{-3mm}\includegraphics[width=7.1cm]{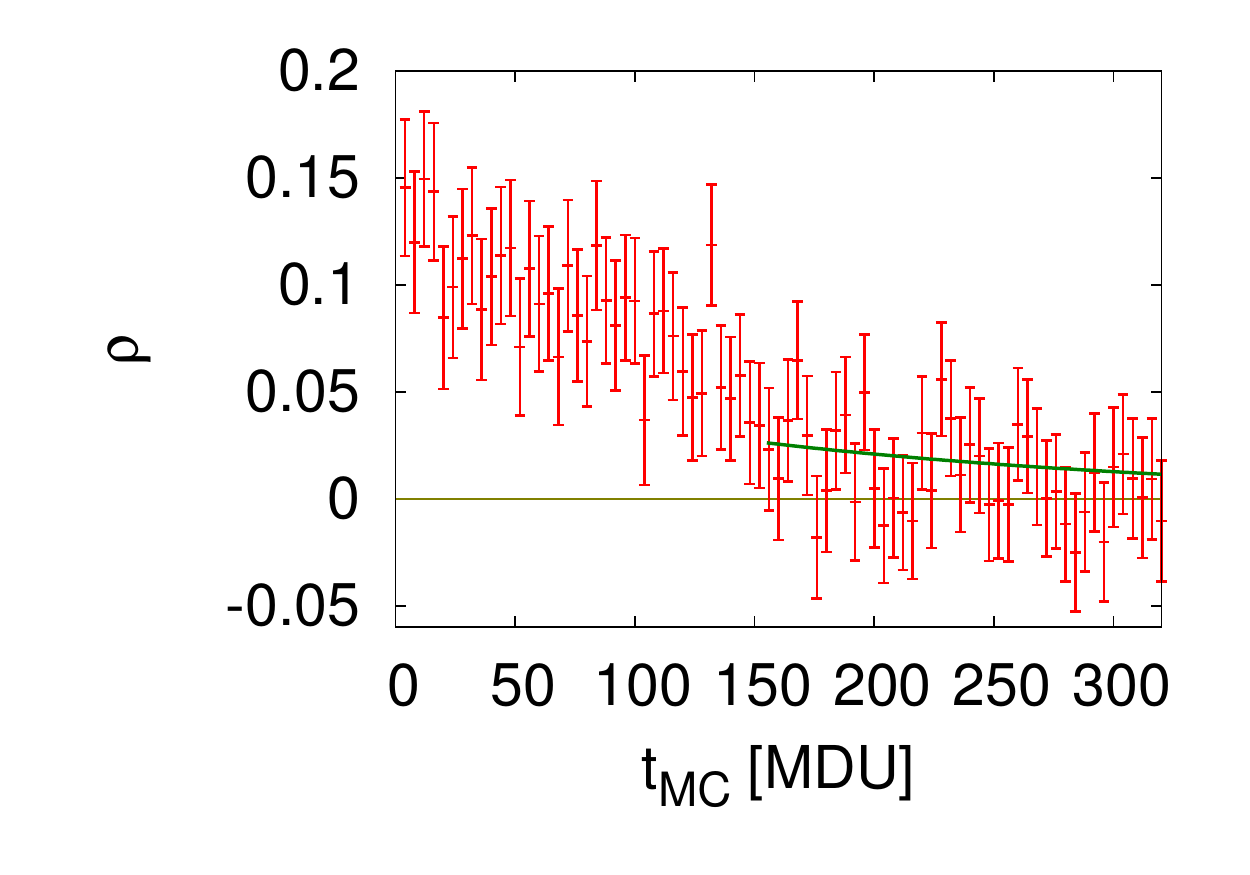}
   \vspace*{-2mm}
\caption{Normalised ($\rho(0)=1$) autocorrelation function of $\fpi$
on CLS ensemble N6 \cite{lat13:Stefano}.
We show the estimate of a contribution of a tail, following
the analysis of \protect\cite{algo:csd}. 
\label{f:rho}
}
\end{wrapfigure}

I would like to
emphasise a further feature of the error analysis of the decay constants. 
We have learnt in 
recent years that one has to be careful concerning the contribution 
of slow modes of the Markov matrix to the autocorrelation function
$\rho_O(\tmc)$ of an observable $O$. Such modes 
contribute as one (or several) slowly decaying exponentials $A \exp(-\tmc/\tauexp)$. Pseudo scalar correlators, 
at least those computed with a noisy estimator
for the source-timeslice spatial average of the correlation function,
show a rather small amplitude $A$ of these potentially dangerous terms. 
I show an example for the aforementioned N6 ensemble in \fig{f:rho}.
This ensemble has a relatively long exponential autocorrelation 
time $\texp\approx 200$ and thus even a small amplitude of 
$A\sim 1/20$ leads to a contribution of order 10 in the integrated
autocorrelation time. Still this is a small $\approx 50\%$ 
part of $\tauint$.
Most of the relevant (and accessible) part of the 
autocorrelation function is dominated by shorter time scales of 
the MC process.

Note that the discussion of autocorrelation issues in QCD simulations
is only semi-quantitative, in particular our estimates of the tails
according to \cite{algo:csd}. However, this is hardly avoidable given
our still limited abilities to generate long MC ensembles. A 
consideration of a possible tail and some estimate  
of the influence of a tail is important exactly because 
statistics is not large enough to just sum up $\rho(\tmc)$ to sufficiently
large $\tmc$.

Despite these problems, my message here is that autocorrelations are a 
minor problem for the decay constants, as long as one is not in the 
desastrous region of parameters, which as far as we know starts
rather close to $a\approx 0.045$~fm~\cite{algo:csd,algo:openbc,Bazavov:2010xr}
for periodic boundary conditions.

\subsection{A scale from the vector correlator}
Over the years an impressive  phenomenological knowledge about the 
vector-meson spectral function has been built up, largely due to the 
efforts to provide a precise number for the hadronic vacuum polarisation contribution to the magnetic moment of the muon,
$g_\mu$~\cite{review:gm2}. David Bernecker and Harvey Meyer 
recently proposed
to build on this knowledge and predict the isovector Euclidean 
time-slice correlator $C_\mathrm{V}(x_0)$ from 
phenomenology~\cite{Bernecker:2011gh}. The
scale $\tau_1$ defined through~\cite{Bernecker:2011gh,lat13:harvey}
\bes
    \left.-x_0  \frac{\rmd}{\rmd x_0}\,\log\,C_\mathrm{V}(x_0)\right|_{x_0=\tau_1} = 3.25
\ees
can then be evaluated from phenomenology to be $\approx0.73$~fm 
in $\nf=2$ QCD. Note that the ``hidden strangeness'' contributions
are removed from the experimental value in order to arrive at a prediction more
closely related to $\nf=2$ QCD. Such a step is obviously not 
done/possible when one fixes the scale through $m_\Omega$ or $f_\pi$.
The particular choice of 3.25 on the right hand side is a 
compromise between a large enough $x_0$ in 
$\frac{\rmd}{\rmd x_0}\log C_\mathrm{V}(x_0)$ beyond the perturbatively dominated
regime with $\frac{\rmd}{\rmd x_0}\log\,C_\mathrm{V}(x_0) \sim 1/x_0$ and to have a good statistical precision. Also 
finite size effecs grow with $x_0$.

An advantage of $\tau_1$ compared to hadron masses is that it is defined
directly at finite $x_0$. But the necessary 
phenomenology is not entirely straight forward. In particular one
has to separate the different isospin 
contributions in the experimental $e^+e^- \to \mathrm{hadrons}$ 
cross section in order to obtain the {\em isovector}  
spectral function or one has to use the full electromagnetic
current which requires quark-line disconnected 
contibutions. 
The method is presently being developed and tested~\cite{lat13:harvey}.

\section{Theory scales}
\label{s:theo}

Already the phenomenology needed for $\tau_1$ is somewhat involved.
We now turn to scales defined in terms of the static quark potential.
Even though a prediction of $r_0\approx 0.49$~fm based on
potential models for $\rm b \bar b\,,\; c \bar c$ spectra~\cite{pot:r0}
has been rather successful, the 
connection of the phenomenological potentials to the static potential
$V(r)$
has never become truely quantitative. I therefore consider both  $r_1$
and $r_0$ as theory scales.

\subsection{$r_0,\,r_1$}
 Their definition is~\cite{pot:r0} 
\bes
    \left. r^2 F(r)\right|_{r=r_c} = c\,,\quad  r_0 \equiv r_{1.65}\,.
\ees
The original motivation for the definition of $r_c$ was an
improvement over the string tension, which was used 
extensively as a reference scale in the pure gauge theory.
The string tension requires a double limit of large time extent of a 
Wilson loop at fixed $r$ and then the limit of large $r$. 
The scales $r_c$ ``only'' require large time extent to extract
the ground state potential. 


Let me discuss a few properties. The force is given by the 
derivative of the potential. When a proper lattice derivative
is used and and the (light-quark and gluon) action
is $\rmO(a)$ improved, the force is $\rmO(a)$ 
improved. This property follows\cite{pot:intermed} from the 
automatic $\rmO(a)$ improvement of static quark actions
\cite{zastat:pap1}. In fact, different static actions with 
a (moderate!) smearing of the gauge field in the static
action can be used~\cite{HYP,stat:action}. In practise, 
apart from unsmeared links, corresponding to the 
Eichten-Hill action for the static quarks, mostly 
the HYP2 action~\cite{stat:action} has been adopted. 
The reason is as follows. The relative errors of 
Wilson loops grow as
\bes
  R_{N/S}^\mathrm{W} &\simas{T\; \mathrm{large}}& K_\mathrm{W}(r)\, \exp\left([\frac{e_1 g_0^2 +\ldots}{a} +\varepsilon(r)]T\right) \,,
  \label{e:rnswils}
\ees
where the mass scale in the exponent diverges in the
continuum limit while 
$\varepsilon(r)$ is finite.  The
divergent term is the self energy of the static quark or,
in a different language,
the perimiter term in the Wilson loop. The coefficient $e_1$ 
depends on the static quark action and the HYP2 variant
has been constructed such that it is rather small. One should keep in mind
that different static actions mean different discretisation effects 
and hence different values for $r_0,r_1$ at finite lattice spacing.

The divergence in the exponent of \eq{e:rnswils} will eventually 
become a serious problem when the 
continuum limit
is approached more and more, but at present lattice spacings,
the generalised eigenvalue method~\cite{alpha:sigma,gevp:pap} combined with 
smearing of the spatial parallel transporter still allows
for sub-percent precision \cite{pot:nf2}. As shown in \fig{f:plateaux}
this method achieves a very early plateau in $V(r)$. 
On the other hand, 
the MILC collaboration adopted $r_1$ as their standard~\cite{pot:r1} since
at smaller $r$ the excited state corrections to the ground state 
decay much faster with $T$, allowing for smaller $T$ and associated
smaller errors, see \fig{f:plateaux}. 
Of course, smaller $r/a$ comes with the price
of generically larger $a^2$ effects. To a certain extent,
kinematical $a^2$ effects can be supressed
by a good definition of the discretised force \cite{pot:r0}
or by a phenomenological subtraction of $a^2$ effects from the 
potential~\cite{pot:r1}. 

Contributions of slow modes to the autocorrelation function of $r_0,r_1$
are less relevant than for $\fpi,\fk$, simply due to the relatively
large variance of the Wilson loops.

\subsection{Scales derived from the gradient flow}
\label{s:flow}
\subsubsection{Gradient flow and definition of $t_0,w_0$.}

The gradient flow is discussed at this conference by Martin L\"uscher~\cite{lat13:martin}. Here we only need the Yang-Mills flow. 
In continuum notation, one introduces
a gauge field $B_\mu(x,t)$ which depends in addition to the space-time coordinates $x$ on a flow time $t$ and
coincides with the quantum gauge field $A_\mu(x)$ at $t=0$. At positive
flow-time, $B_\mu$ is defined through the flow equation~\cite{flow:ML}
\bes
   \frac{\rm d}{{\rm d}t}  B_\mu(x,t) =  D_\nu G_{\nu\mu}(x,t)
   = -{\delta S_{\rm YM}[B] \over \delta B_\mu(x,t)}\,, 
   \quad  B_\mu(x,0) =A_\mu(x)
   \label{e:flow}
\ees
where $D_\nu$ is the covariant derivative in terms of
the gauge field $B_\mu$ and $G_{\mu\nu}=[D_\mu,D_\nu]$.   
As a solution of \eq{e:flow}, $B_\mu$ 
is just a functional of the (true, quantum) gauge field. 
At lowest order in the weak coupling expansion, the flow
equation becomes the heat equation with solution
\bes
    B_{\mu}(x,t) = \int\rmd^4y\; 
                        (4\pi t)^{-2} \rme^{-{(x-y)^2/(4t)}}\;A_\mu(y)
                    +\rmO(g_0^2)\,.
    \label{e:flowpert}                
\ees
We see that the gauge field has been smoothed over a radius 
of $\sqrt{8t}$. 
For $t>0$, correlation functions of this smooth field 
are finite at any Euclidean distance~\cite{flow:LW}, in particular
\bes
   \expe(t) = t^2 \langle E(x,t)\rangle \,,
   \quad E(x,t)\equiv -\frac12 \tr\, G_{\mu\nu}(x,t)G_{\mu\nu}(x,t)
\ees
needs no renormalisation beyond the one 
of gauge coupling and quark masses. Thus, for example in $\nf=2$ QCD,
$f(t / r_0^2, \mpi^2 r_0^2) = \expe(t) $
is a finite function.
The proof of finiteness to all orders of perturbation theory
uses a formulation in terms of a 5-dimensional theory,
adding the extra dimension $t$ (with dimension length$^2$) and a
boundary $t=0$~\cite{flow:LW}. This 5-d formulation is also essential
to perform a Symanzik analysis of cutoff effects~\cite{flow:ferm}
which I turn to in the following section.

First I define the scales~\cite{flow:ML,flow:w0}
\bes
   t_0&:&\;\expe(t_0) =0.3\,,  \label{e:t0} 
   \\
   w_0&:&\;w_0^2 \expe'(w_0^2) =0.3\,,  \label{e:w0}
\ees
where $\expe'(t)=\frac{\rmd}{\rmd t} \expe(t) $. These are truely
theory scales; we need to determine their values in terms
of a phenomenological scale by
a lattice QCD computation, performing the continuum limit.

\begin{wrapfigure}{l}{7.9cm}
   \includegraphics[width=7.9cm]{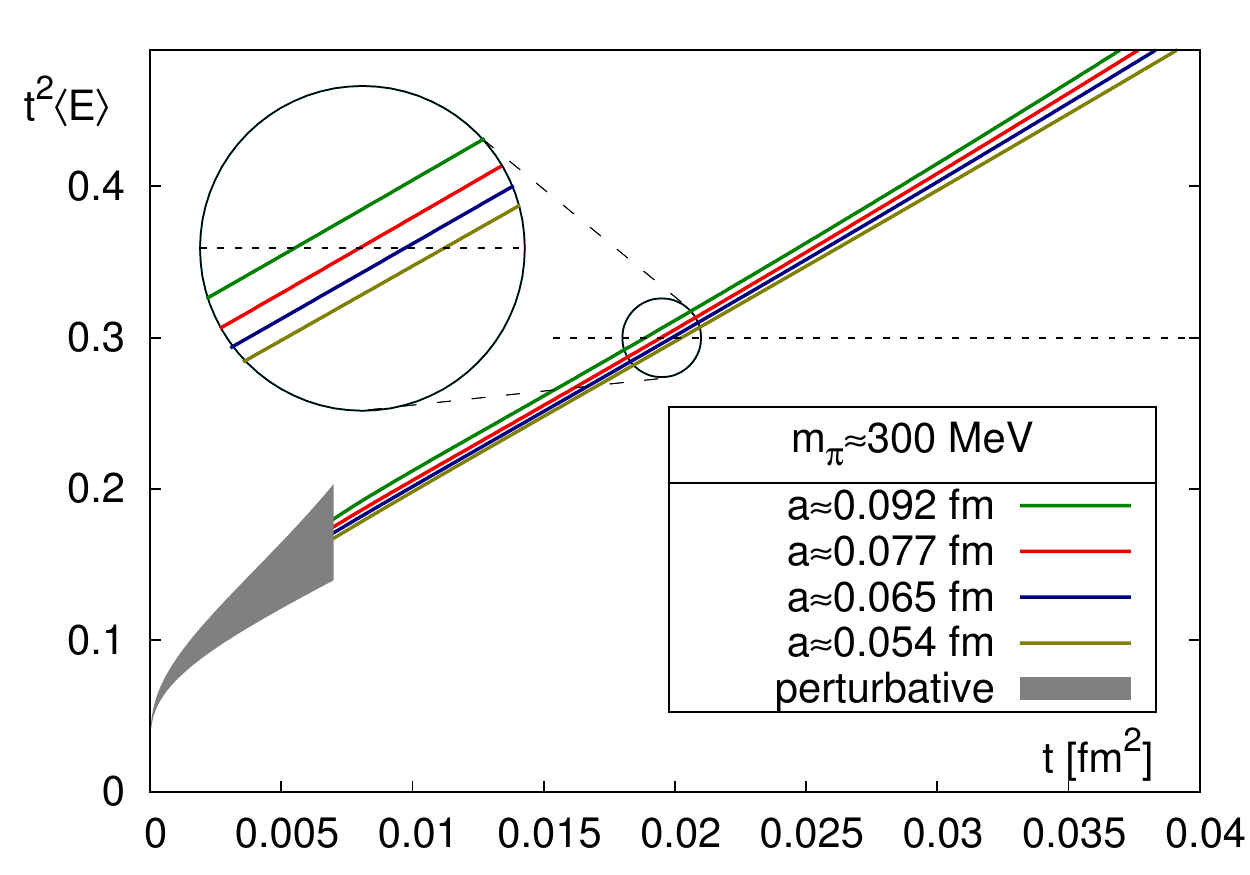}
\caption{$\nf=2+1$ results for $\expe(t)$ from the BMW collaboration.
Graph from \cite{flow:w0}.
\label{f:et}
}
\end{wrapfigure}

\Fig{f:et} shows the shape of $\expe(t)$ computed by the BMW collaboration
in $\nf=2+1$ QCD. In the considered range it is roughly linear in $t$. 
I emphasise that this is a numerical observation and it is approximate.
We have no theoretical reason for this behaviour. 
The strong $t$-dependence of the dimensionless combination 
$\expe(t)$ motivated the definition of $t_0$ by \eq{e:t0}~\cite{flow:ML}. 
Later, the BMW collaboration observed
that, varying the lattice spacing  set by $m_\Omega$, 
there are roughly parallel but somewhat displaced curves. 
This numerical 
observation motivated the definition of $w_0$ in terms of the slope, 
\eq{e:w0} leading to reduced lattice spacing effects compared to $t_0$.
One should bear in mind that parallel lines
in \fig{f:et} turn to crossing ones, when the scale is set 
in a different way which has non-negligible $\rmO(a^2)$ 
cutoff effects relative to $m_\Omega$. 
We turn to a discussion of cutoff effects specific to the gradient flow.

\subsubsection{Cutoff effects}
I start with some heuristics. Expressing quantities 
as integrals in momentum space, 
those dominated by small momenta have smaller cutoff effects than 
those dominated by large momenta, at least as long as no 
accidental cancellations take place. Fourier transforming 
\eq{e:flowpert} yields a gaussian damping of large momenta and
at lowest order of perturbation theory, in the continuum, we obtain
\bes
  \expe(t)  &\sim & g_0^2\, t^2\int_0^\infty {p^3\,\rme^{-2tp^2}} \rmd p
  \label{e:t0int} +\rmO(g_0^4) \\ 
 t \expe'(t) &\sim & g_0^2 t^2\,
    \int_0^\infty {p^3\,(1-t\,p^2)\rme^{-2tp^2}} \rmd p +\rmO(g_0^4) \,.
   \label{e:w0int}
\ees
The integrand of
\eq{e:t0int} is peaked around $p\approx t^{-1/2}$ while in
\eq{e:w0int} it has two peaks below and above that region 
whose integrals cancel. The essential observation is the 
damping of large momenta which renders $\expe$ finite and supresses cutoff effects.  

However, we note that $t_0 \approx (0.15\,\fm)^2$ in QCD. 
Thus with a lattice spacing of $a\approx 0.1\,\fm$, 
the dominating momenta have $ap \approx 0.7$. There, lattice 
momenta $\hat p_\mu^2 = (2/a)^{2} \sin^2(ap_\mu/2)$ differ from
continuum ones by a few percent. In fact, for the standard discretisation
(Wilson flow and plaquette discretisation for $E(x,t)$) 
one obtains 
\bes
    \frac{\expe_\mathrm{lat}(t)}{\expe(t)} 
    = 8(t/a^2)^2 \pi^2 \left\{ \rme^{-4t/a^2}I_0(4t/a^2) \right\}^4 
    = 1 + k_1 \frac{a^2}t + \ldots
    \label{e:e_asq}
\ees
with $k_1\approx 0.13$ and small higher order terms for, say, 
 $a^2/t \leq 1/2$. Similarly,  
$|t{\expe'_\mathrm{lat}(t)}/{\expe(t)}| \approx 0.12 \frac{a^2}t$ 
in the same range, where we normalise to $|\expe(t)|$ since
$\expe'$ vanishes in the continuum limit at this order in the coupling. Obviously such lowest order perturbation theory
estimates are just indicative. 

While these are no particularly large cutoff effects, 
one may be interested in reducing them by Symanzik improvement
and indeed, the Symanzik flow has been used in 
\cite{flow:w0,lat13:brown}.
``Symanzik flow'' refers to the 
flow equation, continuous in $t$, but with the gradient on the 
rhs. of the equation taken as the gradient of the tree-level 
Symanzik improved gauge action, while ``Wilson flow'' refers to the
gradient of the Wilson plaquette action. 
Since the flow equation is a classical equation, without any explicit
coupling constant, it is $\rmO(a^2)$ improved by using the
{\em tree level} Symanzik improved action. However, this is not at all
sufficient for full $\rmO(a^2)$ improvement. For
full improvement, the following ingredients would be necessary:
\bi
\setlength{\itemsep}{0pt}
\item Symanzik flow, i.e. improvement of the (classical) flow equation.
\item Improvement (classical) for the discretisation of $E(x,t)$ .
\item Improvement (with $g_0^2$ dependent coefficients) at the $t=0$
    boundary.
\item Improvement (with $g_0^2$ dependent coefficients) of the 4-d action.
\ei
This discussion uses the 5-d path integral (with Lagrange multiplier fields) \cite{flow:LW,flow:ferm} 
and improvement of a theory with boundaries as previously 
developed for the Schr\"odinger functional. While the first two 
requirements are "easily" done by switching to the Symanzik flow
and the Symanzik action density for $E(x,t)$, there are many terms
for the other two.
For example at the boundary one can have
$t$-derivatives. With a dimension four field $\cal O$, the term
$\left. a^2 \int \rmd^4 x \frac{\rmd}{\rmd t} {\cal O}(x,t) \right|_{t=0}$ 
is dimensionless and has to be considered as a term in the 
improved action. Systematic, non-perturbative, $\rmO(a^2)$ improvement seems out of reach! I discuss this here because after all the Symanzik flow 
is used in paractical computations.

Let me indicate the degree of improvement {\em of the flow} by 
(flow,discr), where flow is either ``Sym'' for the Symanzik flow or 
``Wils'' the Wilson flow and discr is ``plaq'' for the plaquette 
discretisation of $E(x,t)$ or ``clov'' for the symmetric 
definition~\cite{flow:ML} formed from the clover $G_{\mu\nu}$.
Boundary terms have so far not been considered. 
I therefore do not include them in my notation. 
Furthermore the
degree of improvement of the 4-d action is indicated separately. 

Incomplete improvement may be worse than no
improvement. An example has been seen by A. Ramos in the 
leading order perturbative computation
of $\expe(x,t)$ in a Schr\"odinger functional setting, with 
(Wils,clov) once evaluated with the tree level improved 
4-d action and once with the Wilson action. 
Improving just the 4-d gauge action leads to a
factor $\approx2$  increase of the cutoff effects in this 
case at leading order in the coupling.

I note that in Martin L\"uscher's original pure gauge theory demonstration
the difference of (Wils,clov) and (Wils,plaq) amounted to $\approx 6$\%
in $t_0/r_0^2$ at $a=0.1\fm$. This is the same magnitude
as the estimate \eq{e:e_asq}.

Given the difficulties in systematically also removing 
$a^2$ terms, I think it is worth to consider to define a scale $t_1$
simply at somewhat larger flow time. In particular,
\bes
   \expe(t_1)=2/3\, \label{e:t1}
\ees   
seems a good choice.
At $t_1$ the kinematical cutoff effects \eq{e:e_asq} are 
about a factor two suppressed (and better approximated 
by just $k_1a^2/t$) and the same would be the case for an
analogous $w_1$. Obviously, $t_1$ will, however, 
be more sensitive to finite volume effects, which needs to be 
investigated.

\subsubsection{Precision}
A tremendous advantage of $t_0$ and $w_0$ is the high statistical precision.
The variance of these quantities is very small and I note that due
to the finiteness of correlation functions at all distances, it 
also remains finite in the continuum limit. On the other hand,
at finite $t$
the autocorrelations are very much enhanced, see \cite{lat13:mattia,lat13:brown,algo:openbc}.
Despite this, the variance is so small that with a run of length
$t_\mathrm{MC} / \tauint(t_0) \approx t_\mathrm{MC} / \tauexp = 20$
one achieves a precision of $\sqrt{t_0}/a$ at the level of around
one per mille on a lattice of $3^3\times6\,\fm^4$. On larger lattices
self averaging improves the precision further. 

A second advantage of these scales is that $\expe(t)$ is 
obtained as a
straight expectation value. One does not need to extract the large
time decay of some correlation function as it is the case for
masses or $r_0,r_1$. Since the integration of the flow equation 
can be done with very high numerical precision,
there is no systematic error -- apart from the 
unavoidable discretisation and finite volume effects. 

And finally, there is a weak quark mass dependence. 
The small quark mass expansion 
of expectation values of ``local'' 
fields, at fixed $t$,  can be computed in chiral perturbation
theory \cite{lat13:martin}. 
The NNLO formula for $\expe(t)$ has been worked out 
by O.~B\"ar and M.~Golterman \cite{flow:t0chpt}, with the interesting 
result that at and including order $\mpi^2$ there are no 
non-analytic terms. This is in qualtitative agreement 
with the very linear mass-dependence seen in the chiral extrapolations,
for example \fig{f:t0mass}, below.

\section{Status}
\label{s:status} 

I now want to give an impression how well we know the various 
scales at present. I will rather uncritically cite numbers from
the literature and I would like to apologise that it is impossible
to do justice to all computations. I tried to take into account
the more recent numbers from the larger simulations.

\subsection{$m_\Omega$ and $m_\prot$}
The masses of the Omega and the proton are experimentally
known, so the relevant question is what kind of precision 
we are able to achieve in lattice computations. 
I neglect here the issue of the systematic uncertainty in 
quark mass extra/interpolations, even though this is rather 
relevant. I just note that precisions quoted for 
$m_\Omega$
(usually after the extrapolation to the physical point) are
$\approx 0.5$ \% \cite{Aoki:2009ix}, 
$\approx 1$ \% \cite{Arthur:2012opa}, $\approx 1$\% \cite{momega:bmw}, 
$\approx 5$ \% \cite{momega:cls}. Obviously, such numbers depend on 
the statistics. Nowadays it is typically around 1000 to 10000 
molecular dynamics units. But they also depend on the 
starting value of a fit to a plateau, with an associated systematic
error which is not so easily quantified.

\subsection{$r_0$ and $r_1$}

\begin{table}[t!]
\begin{tabular}{lll lll | llllll } 
    \multicolumn{3}{c}{Wilson, $\Nf=2$} &
    \multicolumn{3}{c}{tmQCD,  $\Nf=2$} & 
    \multicolumn{4}{c}{$\Nf>2$} 
    \\
     $r_0$[fm] &  \multicolumn{2}{c}{from} & 
     $r_0$[fm] & \multicolumn{2}{c}{from} &
     $\Nf$ & $r_0$[fm] & $r_1$[fm] & \multicolumn{2}{c}{from} 
     \\[0.5ex] \hline \\[-1.5ex]
   $    0.503 (  10 ) $& $\fk$ & \cite{alpha:lambdanf2} &      
   $    0.438 (  14 ) $& $\fk$ & \cite{Blossier:2009bx} &
   2+1 & $ 0.466(4)^a$ & $ 0.313(2)$ & div. & 
   \cite{Davies:2009tsa}
   \\
   $    0.491 (  ~6 )^c $& $\fk$ & \cite{lat13:Stefano} &&&&
      2+1 &  & $ 0.321(5)$ & $\Upsilon$ & \cite{Follana:2007uv}
\
   \\
   $    0.485 (  ~9 )^c $& $\fpi$ & \cite{lat13:Stefano}&    
   $    0.420 (  20 ) $& $\fpi$ & \cite{Baron:2009wt} 
   &
   2+1 & $0.470(4)$ & $0.311(2)$ & $\fpi$ & 
   \cite{Bazavov:2010hj,Bazavov:2011nk}
\\  
    $0.501(15)^b$    & $m_{\prot}$ & \cite{Bali:2012qs} &
   $    0.465 (  16 ) $& $m_{\prot}$ & \cite{Alexandrou:2009qu} 
   &   2+1 & $    0.492 (10)^b $ & & $m_\Omega$ & \cite{Aoki:2009ix} \\ 
   $    0.471 (  17) $& $m_\Omega$ & \cite{momega:cls} &
     &&&
   2+1 & $    0.480 (  11 ) $ & $0.323(9)$ & $m_\Omega$ & \cite{Arthur:2012opa}\\ 
   &&& &&&
   2+1+1 & & 0.311(3) & $\fpi$ & \cite{Dowdall:2013rya} 
\\
\end{tabular}
\footnotesize
$^a$ with $r_0/r_1$ and $r_1/a$ from \cite{Bazavov:2009bb} 
\hspace*{2cm} $^c$ preliminary, at this conference
\\
$^b$ no continuum extrapolation
\caption{Values for $r_0,r_1$.  Column ``from'' shows the phenomenological scale 
used. All results except for those marked with $^b$ have been obtained by a
continuum extrapolation. 
\label{t:r0}}
\end{table}

In large scale simulations, 
$r_1/a$ has been determined by MILC, HPQCD, RBC/UKQCD and
HOTQCD; it has been converted to physical units for $\nf=2+1$ 
using a variety of phenomenology scales, see \tab{t:r0}. 
Also modern determinations of $r_0$ are listed in that table.

There is still a considerable spread, 
most notably in the $\nf=2$ theory for $r_0$ between the ALPHA and ETM
collaborations. As the numbers refer mostly to the continuum limit
they should agree.
In \sect{s:problems} I return to the worrysome difference seen 
in \tab{t:r0}.

For $\nf>2$ all determinations performing a continuum
extrapolation are compatible with 
\bes
    r_0 = 0.472(5)\,\fm\,,\quad r_1=0.312(3)\,\fm\,.
\ees
I refrain from performing a weighted average of the numbers in
\tab{t:r0} for the following reasons. 
Systematic errors, e.g. due to chiral extrapolations, may be relevant.  
More importantly, these numbers are entirely dependent on 
a single set of simulations / action, the
 MILC configurations with rooted staggered fermions. 
 There is only a weak (due to the much larger error)
 cross check by the RBC/UKQCD result $r_0=0.480(11)\,\fm$,  
 $r_1=0.323(9)\,\fm$~\cite{Arthur:2012opa}.

\subsection{$t_0$ and $w_0$}
\begin{table}[b!]
\onecol{0.6\textwidth}{
\begin{tabular}{llll} 
    $\nf$ & $\sqrt{t_0}$~[fm] & $w_0$~[fm]  & from\\[0.5ex] \hline \\[-1ex]
       0 &$   0.1638 (  10 ) $&$   0.1670 (  10 ) $& $r_0=0.49\,\fm$ \cite{lat13:mattia,flow:ML} \\ 
   2 &$   0.1539 (  12 ) $&$   0.1760 (  13 ) $& $\fk$ \cite{lat13:mattia,lat13:Stefano} \\ 
   3 &$   0.153~(7)         $&$   0.179~(6)         $& $m_\prot$\cite{lat13:roger} \\ 
   3 &$   0.1465 (  25 ) $&$   0.1755 (  18 ) $& $m_\Omega$ \cite{flow:w0} \\ 
   4 &$   0.1420 (   8 ) $&$   0.1715 (   9 ) $& $\fpi$ \cite{Dowdall:2013rya}\\ 
   4 &$                  $&$   0.1712 (   6 ) $& $\fpi$ \cite{lat13:brown} \\ 
 \\[-1ex]
    \hline \\[-1ex]
\end{tabular}
}
\hfill
\onecol{0.30\textwidth}{\vspace*{-22.48mm}
\includegraphics[width=45mm,height=47.0mm]{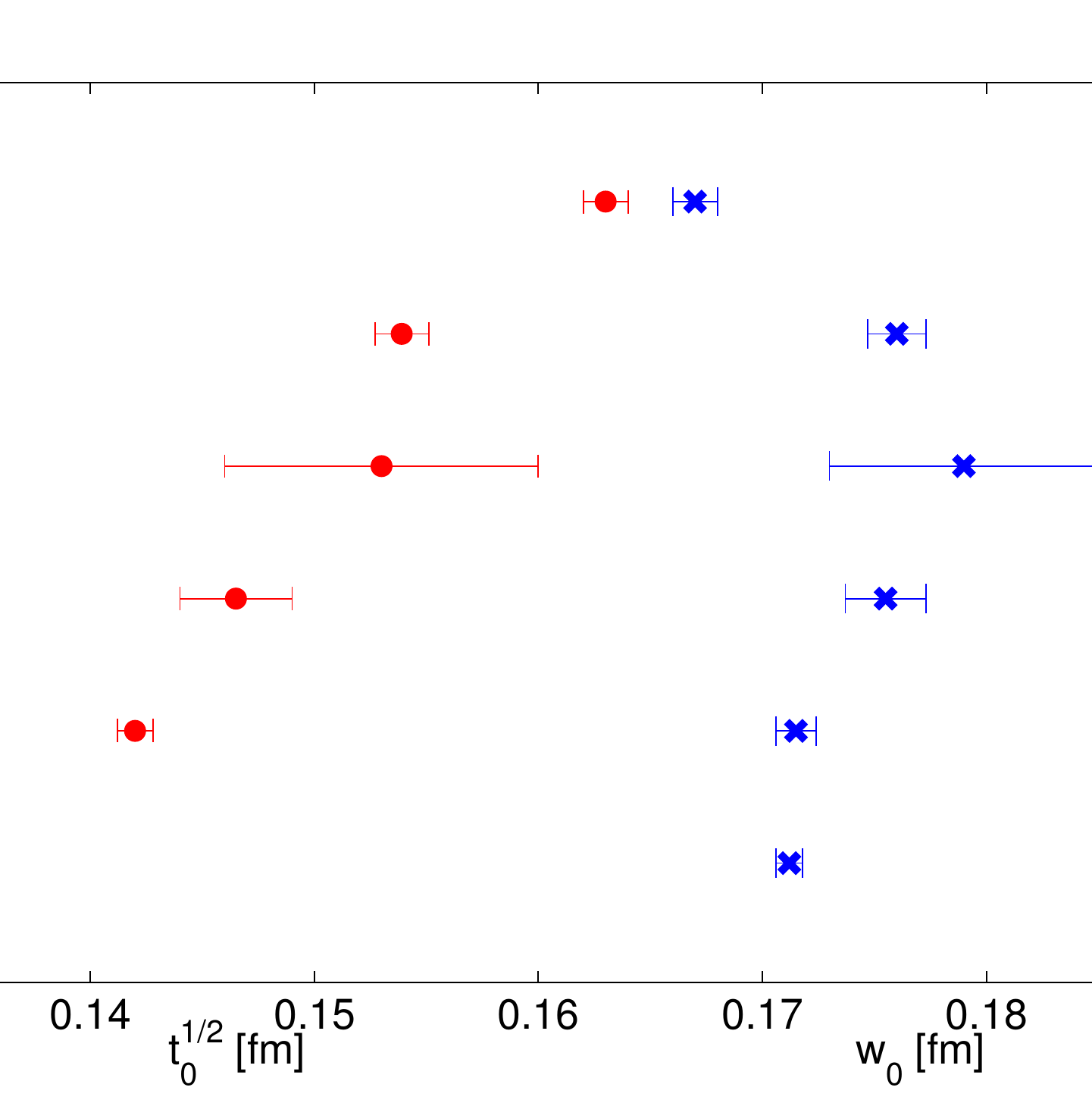}
}
\caption{Scales from the gradient flow. Note that these depend in principle
on the phenomenological scale they are determined from. 
\label{t:scales}
}
\end{table}
\begin{figure}[t!]
\centering
   \includegraphics[width=.495\textwidth]{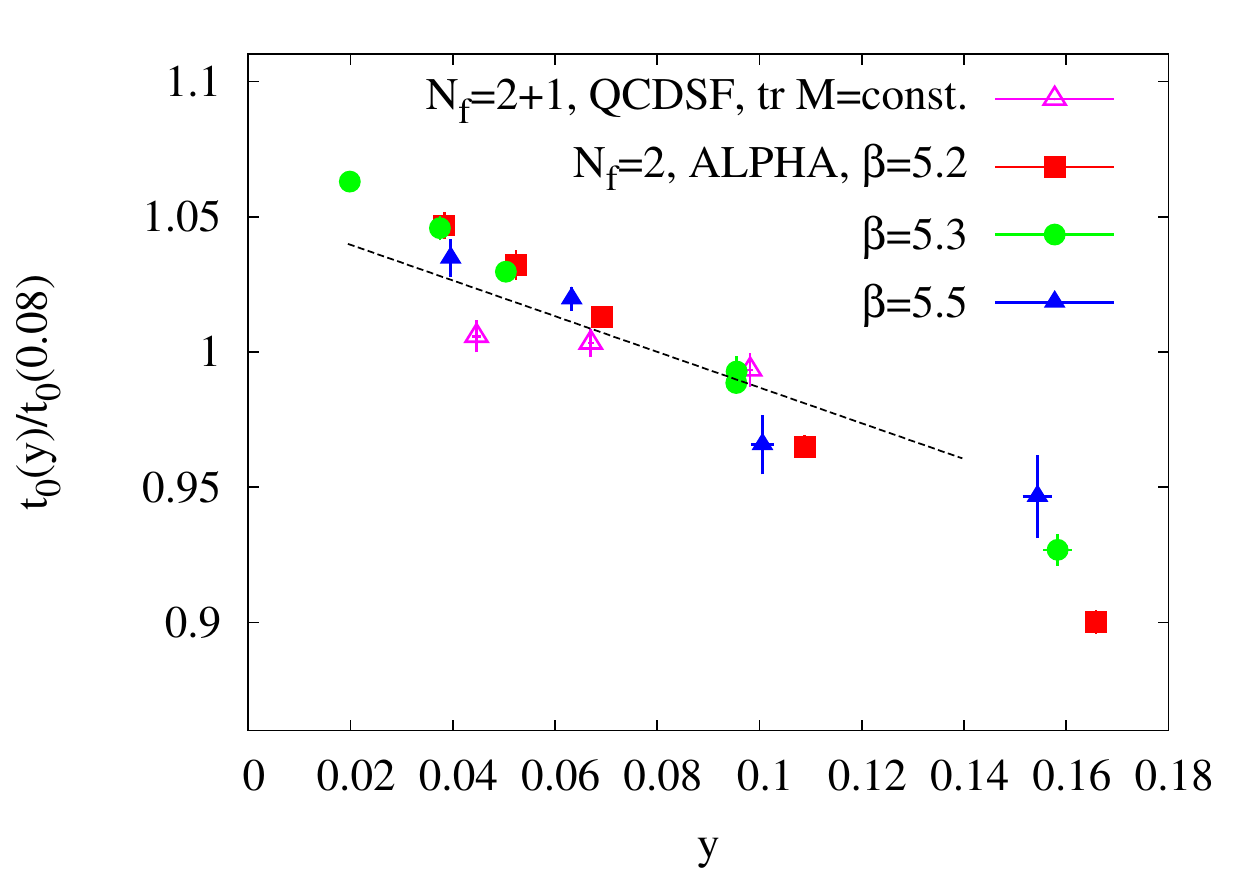}
   \includegraphics[width=.495\textwidth]{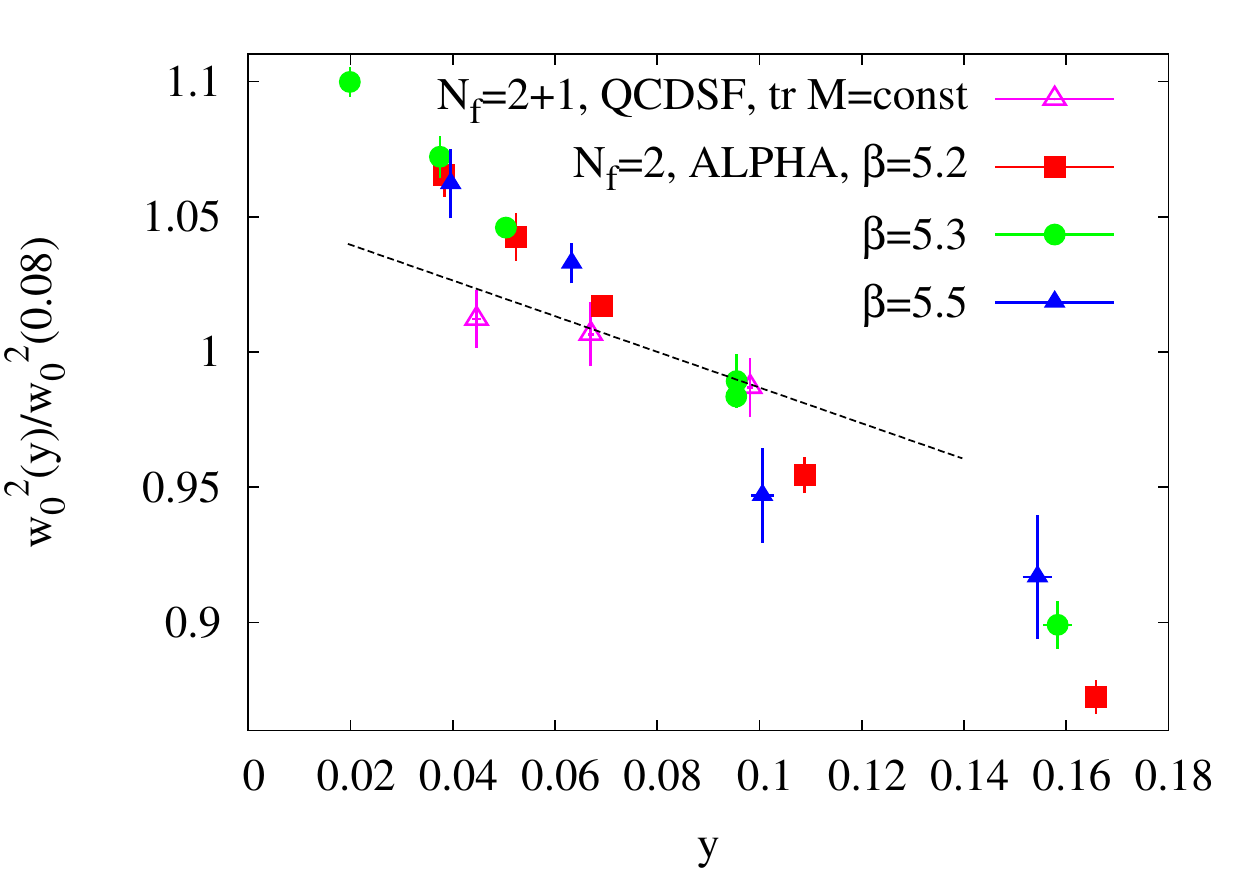}
\caption{Mass dependence of the flow scales. We consider the functions
$t_0(y)$, $w_0(y)$ with the dimensionless measure $y=t_0 \mpi^2$ 
for the quark mass and normalised at the point $y=0.08$. 
Filled symbols are for $\nf=2$ CLS lattices with 
$a=0.045\,\fm \ldots 0.07\,\fm$ \cite{lat13:mattia}
and open symbols are preliminary results
by QCDSF for $\nf=2+1$ along a trajectory $\tr m_\quark$=constant \cite{lat13:roger}.
As a reference the dotted line correspond 
approximately to the behaviour $r_0^2$.
\label{f:t0mass}
}
\end{figure}

Clearly, the newcomers $t_0$ and $w_0$ are most interesting.
I discuss them in some more detail. In \fig{f:t0mass} 
I show the mass dependence for $\nf=2$ and with open symbols 
the dependence in the 2+1 theory along a trajectory with constant
trace of the mass matrix. The results are normalised at an 
intermediate mass point defined by the variable $y=t_0 \mpi^2$, 
where $t_0$ is the 
mass-dependent quantity. The figures show first of all excellent scaling
of the quark mass dependence of the $\nf=2$ results with an
$\rmO(a)$ improved action. Second a remarkably linear 
dependence is seen and third the comparison to the dotted line
referring to $r_0^2(y)/r_0^2(0.08)$ 
indicates that $t_0$ has a somewhat stronger quark mass dependence than 
$r_0^2$ while $w_0^2$ has a significantly stronger one.
As expected (see e.g. \cite{flow:t0chpt}, 
the mass dependence along the trajectory $\tr m_\mathrm{quark}$=constant  
is quite a bit weaker. Rough estimates of the slopes are reported in \tab{t:slopes} and $t_0,w_0$ in physical units in \tab{t:scales}.
 
I have explained earlier that discretisation errors are 
a relevant, but also a subtle subject. General statements
are difficult because
discretisation errors are only meaningful after the scale 
has been set, or equivalently for dimensionless ratios. 
The same is true for the $\nf$-dependence.

\begin{table}[ht!]
\centering
\begin{tabular}{llll | lllllll} 
    $\nf$ & $\sm_{t_0}$ &  $\sm_{w_0^2}$ & Ref. & &
    $\nf$ & $\sa_{t_0/Q}$ &  $\sa_{w_0^2/Q}$ & $Q$ & flow & Ref. 
   \\[1ex] \hline &&&&&&&&&\\[-1ex]
         &    &   &  & & 0 & -1\% & -3\% & $r_0^2$ & Wils & \cite{flow:ML} 
   \\
     2    &  -12\% & -20\% & \cite{lat13:mattia} &&
     2    &   -8\% & -19\% & $r_0^2$ &  Wils & \cite{lat13:mattia} 
\\
     2+1    &       & -18\% & \cite{flow:w0} &&
     2+1    & -19\% & $\approx 0$  & $m_\Omega^{-2}$ &  Sym &\cite{flow:w0} 
\\
     2+1+1    &    & -13\% & \cite{lat13:brown} &&
     2+1+1    &     & $\approx 0$  & $\fpi^{-2}$ & Sym & \cite{lat13:brown} 
\\
        &    & &  &&
     2+1+1    &  $ 7\%$  & $\approx 0$  & $\fpi^{-2}$ & Wils & \cite{flow:hpqcd,lat13:Dowdall} 
\\
\end{tabular}
\caption{Slopes with respect to the mass, $\sm_R$, \protect\eq{e:sm} and
    with respect to the lattice spacing,   
    $\sa_R = {[R]_{a=0.1\fm} /[R]_{a=0}} -1
    $. Note that \cite{lat13:Dowdall} states that the 
    Wilson flow is used.      
\label{t:slopes}
}
\end{table}

\subsection{Comparison of scales and $\nf$-dependence.} 

\begin{figure}[b!]
\subfigure[$t_0/r_0^2$]{
\centering
\includegraphics[width=0.3\textwidth]{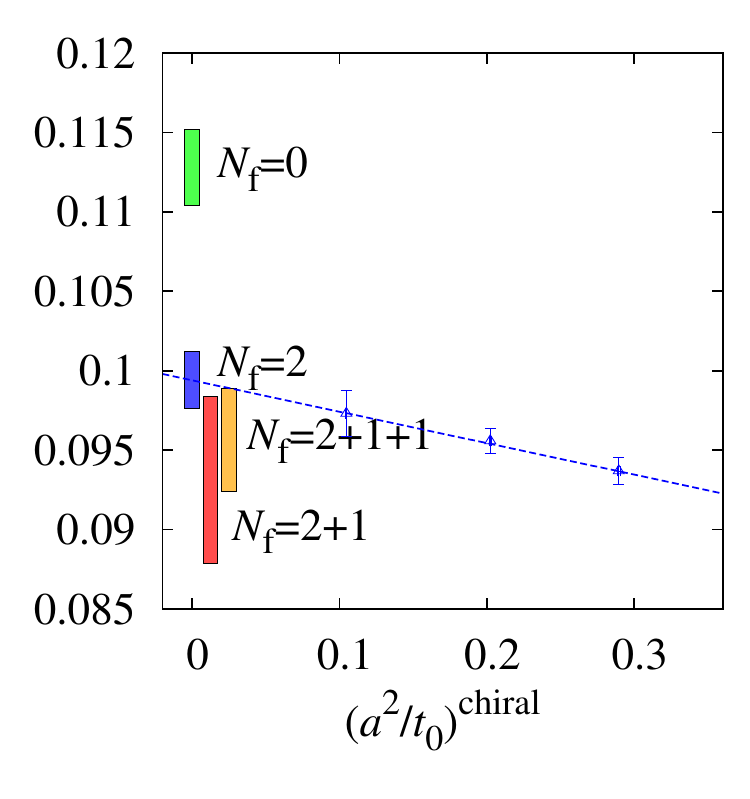}
}
\subfigure[$t_0/w_0^2$]{
\centering
\includegraphics[width=0.3\textwidth]{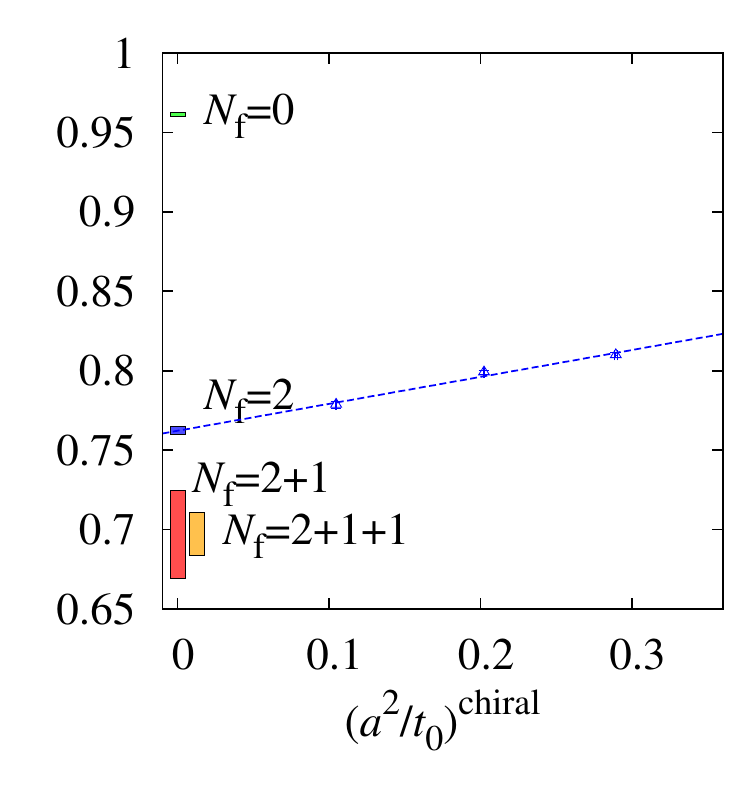}
}
\subfigure[$w_0^2/r_0^2$]{
\centering
\includegraphics[width=0.3\textwidth]{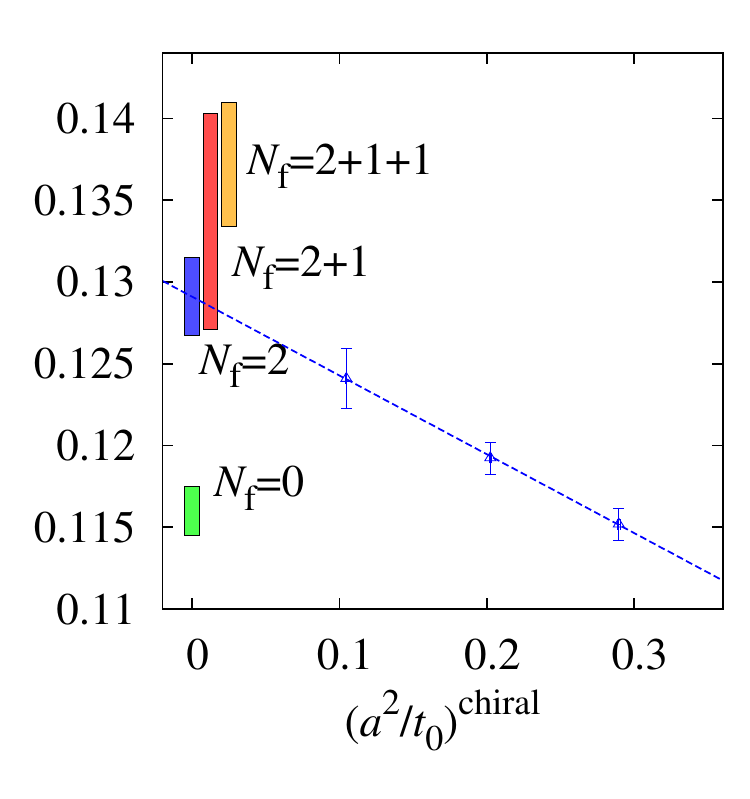}
}
\caption{Continuum extrapolation and flavour number dependence of
ratios of scales. Taken from \cite{lat13:mattia}.
}
\label{f:ratios}
\end{figure}


In \fig{f:ratios}, taken from \cite{lat13:mattia} I show the
approach to the limit $a\rightarrow0$ of ratios of 
the scales $t_0,w_0^2,r_0^2$.
Here  $t_0/r_0^2$ has the smallest discretisation effects,
while in \cite{flow:w0}, BMW reports small $a$-effects for $w_0\,m_\Omega$,
compare for the slopes $S^a$ in \tab{t:slopes}. Whether 
$t_0$ or $w_0$ have smaller cutoff effects depends on
the details of the discretisation and/or the reference scale.
General statements are not possible.

The Figure also shows a comparison with results for 
different $\nf$. 
Here \cite{flow:w0,flow:ML,lat13:brown,Arthur:2012opa,Bazavov:2010hj,Bazavov:2011nk,Davies:2009tsa}
enter, see  \cite{lat13:mattia}.
The ratios demonstrate that the 
$\Nf=0$ and the $\Nf=2$ theories differ quite strongly for these
purely gluonic infrared-dominated, non-perturbative, observables.  
The effects of the heavier quarks,
strange and charm, appear to be less pronounced, but still noticeable.  
Of course, for a very heavy quark, decoupling is expected in such ratios. 
The charm quark may be heavy enough for decoupling to apply 
semiquantitatively. 

A similar $\nf$-dependence is present for the ratio $r_1/r_0$:
It is about $0.66$ for $\nf=2+1$ \cite{Bazavov:2011nk,Arthur:2012opa}, 
for $\nf=2$ it has not directly been computed 
but from \cite{pot:nf2,lat11:bjoern} one can read off 
$r_1/r_0\approx 0.67$ and finally for $\nf=0$ the results 
of \cite{pot:intermed} and others imply $r_1/r_0 = 0.73$. 

At least part of the theory scales in physical units 
have to depend significantly on the number of flavours
since an $\Nf$-dependence is present in the ratios.
Note that in \tab{t:scales} the phenomenological 
reference scale(s) are mostly the light pseudo scalar decay constants,
our best phenomenological scales at present.
So mostly we are looking at a common reference scale.

\begin{figure}[t!]

   \centering
   \includegraphics[width=.45\textwidth]{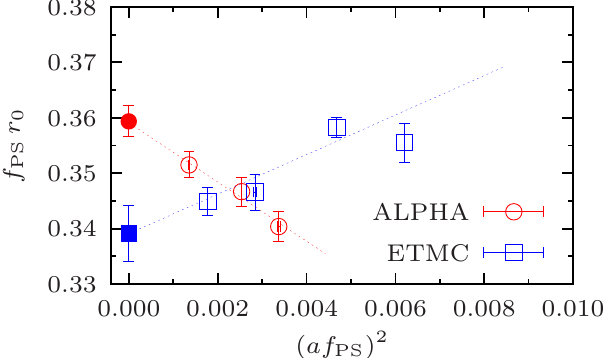}
   \includegraphics[width=.45\textwidth]{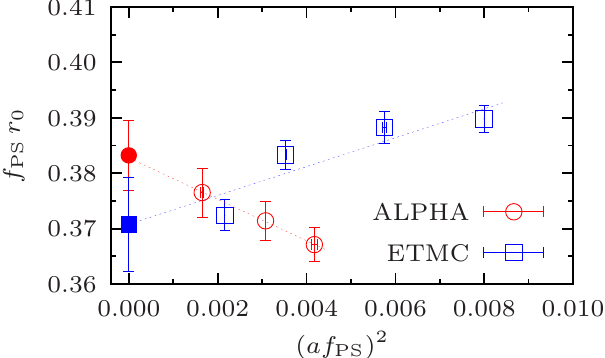}
\caption{ Continuum limit of $f_\pi r_0$ interpolated to $(\mpi*r_0)^2=0.64$ (left)
and to  $(\mpi*r_0)^2=1.128$ (right) for $\nf=2$.
\label{f:etmcalpha}
}
\end{figure}

\section{Open questions}
The considered theory scales are not listed in the particle data book.
We have to determine them  without prior knowledge. They therefore
also constitute a good cross-check of our calculations as we
should agree on their values, at least given the same $\nf$
and experimental input scales. 
Unfortuantely for $\nf=2$ the agreement 
on $r_0$ is not very good.

\label{s:problems}
I therefore briefly discuss the differences in the determinations
by the ALPHA and ETM collaborations. Comparing $r_0$ at the physical point,
as is done in \tab{t:r0}, 
the origin of differences may be due to various sources, where, barring the 
possibility of actual mistakes, two particular ones come to mind.
The first is the chiral extrapolation and the second is finite size effects.
In order to see whether prior to any extrapolation, the two computations/discretisations do agree, G. Herdoiza performed an
interpolation of dimensionless combinations $r_0\fpi$ to two different
reference pion masses, defined by $\mpi^2 r_0^2=0.640, 1.128$. 
S. Lottini and B. Leder added the ALPHA collaboration points with a proper correlated error analysis. The surprising outcome, \fig{f:etmcalpha}, is
that the data actually agree reasonably well at intermediate lattice spacings, 
but the standard continuum extrapolations cross and lead 
to a few sigma difference at
$a=0$ at the smaller pion mass. My attempt to understand this in terms
of the finite $a$ chiral perturbation theory formulae for tmQCD of 
Oliver B\"ar \cite{Wchpt:fpitmqcd} failed. 
In fact the formulae suggest larger cutoff effects 
than the data show, when $m_{\pi^+}^2 - m_{\pi^0}^2$
from \cite{Herdoiza:2013sla} is inserted. Maybe this suggests that the 
mass-splitting is actually smaller.
In any case, the differences remain unexplained at present.
As a speculation I add that there would be very good agreement
if one just took the ETMC data at larger lattice spacings together
with a very flat continuum extrapolation. 

More work on this issue seems warrented. At least the 
analysis should be repeated with $r_0 \to \sqrt{t_0}$
in order to remove the possibility that there is something 
wrong with a determination of $r_0$. 
The present status provokes the questions: {\bf How well do we control our computations? Is our data ready for a continuum extrapolation?}

\section{Conclusions}
This talk was scheduled at a time of transition. The 
all-time favorite $r_0,r_1$ are being replaced by scales derived from the gradient flow. The advantages of the latter are predominantly:
\bi
\item There is an excellent statistical precision. Just 20 independent 
    configurations lead to around per mille precision in the 
    lattice spacing (of course the error is rather uncertain with only 20)
    when the lattice volume is reasonably large.
\item Systematic errors due to excited state contributions are 
    entirely absent in $\expe(t)$ and consequently, 
    as remarked in the question session at the conference, 
    it is easy to compute $t_0,w_0$. There are no pitfalls 
    in determining such scales.    
\ei
In my opinion the second point is very important. 
The possible danger of a misidentified plateau is  
not immediately evident in \fig{f:plateaux}. But 
for the example of the potential at large distances,
where string breaking occurs,
it has been seen that one has to be very careful in the 
selection of the right correlation function to find
the correct ground state energy. More trivially, one
may easily select a plateau value which is a sigma or two away from
the true one because the errors at larger time mask 
the continuing variation of the effective mass. Only with
plateaux as long as they are seen in the pseudo scalar 
sector one feels really safe.

The numbers listed in \tab{t:scales} do largely come from 
a preliminary analysis for this conference. They should better be reviewed 
again in one or two years time.

I think that it may be an improvement to define
a scale at lower momenta; $t_1$, \eq{e:t1} seems promising.
It will be less sensitive to the details of the
definition of the flow equation (concerning $a^2$ terms).
Furthermore, a bit of reflection on the reported size of 
mass- and $\nf$-dependence leads to the 
conclusion that both of these will be rather small for $t_1$.
These advantages should be weighed in comparison to an expected increase
in finite size effects which remain to be evaluated.


Finally I hope that the community will not leave the problem of 
\sect{s:problems} behind, but will solve it soon. Interesting 
physics results have been obtained with the scale from $\fpi$ and $r_0$;
some of them are certainly affected.

\acknowledgments
I would like to thank Martin L\"uscher for very useful discussions 
before and after my presentation as well as for  
sharing his $\Nf=0$ data for the flow observables. I profited 
from work of the ALPHA collaboration, in particular by Mattia Bruno and 
Stefano Lottini. I thank Oliver B\"ar, Nathan Brown, Michele Della Morte, 
Albert Deuzeman, Gregorio Herdoiza, Georg von Hippel, 
Roger Horsley, Benjamin Jaeger, 
Bj\"orn Leder, Harvey Meyer, Alberto Ramos, Stefan Schaefer, 
Carsten Urbach 
for early communication of not yet
published material and stimulating discussions. Many thanks go to 
Tom de Grand for relevant questions on the talk and Ulli Wolff for
a critical reading of the manuscript.
I acknowledge support by the Deutsche Forschungsgemeinschaft (SFB/TR~9)
 and the European Community (grant 283826, HadronPhysics3). 

\bibliography{latticen}
\bibliographystyle{JHEP}

\end{document}